\documentclass[sigconf]{acmart}
\usepackage{array}

\usepackage{multirow}

\newcommand{\gdh}[1]{{\textcolor{black}{#1}}}
\newcommand{\yzm}[1]{{\textcolor{black}{#1}}}

\newcommand{\ghn}[1]{{\textcolor{black}{#1}}}

\AtBeginDocument{%
  }

\setcopyright{acmlicensed}
\copyrightyear{2018}
\acmYear{2018}
\acmDOI{XXXXXXX.XXXXXXX}

\acmConference[Conference acronym 'XX]{Make sure to enter the correct
  conference title from your rights confirmation emai}{June 03--05,
  2018}{Woodstock, NY}
\acmISBN{978-1-4503-XXXX-X/18/06}

\begin{document}

\title{MLoRA: Multi-Domain Low-Rank Adaptive Network for Click-Through Rate Prediction}
\author{Zhiming Yang}
\authornote{These authors have contributed equally to this work.}
\email{yangzhiming@mail.nwpu.edu.cn}
\affiliation{%
  \institution{Northwestern Polytechnical University}
  \city{Xi'an}
  \country{China}
}

\author{Haining Gao}
\authornotemark[1]
\email{gaohaining.ghn@alibaba-inc.com}
\affiliation{%
  \institution{Alibaba Group}
  \city{Hangzhou}
  \country{China}
}

\author{Dehong Gao}
\authornotemark[1]
\email{dehong.gdh@nwpu.edu.cn}
\affiliation{%
  \institution{Northwestern Polytechnical University}
  \city{Xi'an}
  \country{China}
}

\author{Luwei Yang}
\authornote{Corresponding authors.}
\email{luwei.ylw@alibaba-inc.com}
\affiliation{%
  \institution{Alibaba Group}
  \city{Hangzhou}
  \country{China}
}

\author{Libin Yang}
\authornotemark[2]
\email{{libiny}@nwpu.edu.cn}
\affiliation{%
  \institution{Northwestern Polytechnical University}
  \city{Xi'an}
  \country{China}
}

\author{Xiaoyan Cai}
\email{xiaoyanc@nwpu.edu.cn}
\affiliation{%
  \institution{Northwestern Polytechnical University}
  \city{Xi'an}
  \country{China}
}

\author{Wei Ning}
\email{wei.ningw@alibaba-inc.com}
\affiliation{%
  \institution{Alibaba Group}
  \city{Hangzhou}
  \country{China}
}

\author{Guannan Zhang}
\email{zgn138592@alibaba-inc.com}
\affiliation{%
  \institution{Alibaba Group}
  \city{Hangzhou}
  \country{China}
}

\renewcommand{\shortauthors}{Trovato et al.}

\begin{abstract}
Click-through rate (CTR) prediction is one of the fundamental tasks in the industry, especially in e-commerce, social media, and streaming media. It directly impacts website revenues, user satisfaction, and user retention. 
However, real-world production platforms often encompass various domains to cater for diverse customer needs. 
Traditional CTR prediction models struggle in multi-domain recommendation scenarios, facing challenges of data sparsity and disparate data distributions across domains.
Existing multi-domain recommendation approaches introduce specific-domain modules for each domain, which partially address these issues but often significantly increase model parameters and \ghn{lead to} insufficient training. 
In this paper, we propose a Multi-domain Low-Rank Adaptive network (MLoRA) for CTR prediction, where we introduce a specialized LoRA module for each domain.
This approach enhances the model's performance in multi-domain CTR prediction tasks and is able to be applied to various deep-learning models.
We evaluate the proposed method on several multi-domain datasets. 
Experimental results demonstrate \gdh{our MLoRA approach achieves a significant improvement} compared with state-of-the-art baselines.
\yzm{Furthermore, we deploy it in the production environment of the Alibaba.COM~\footnote{\href{https://www.alibaba.com/}{https://www.alibaba.com/}}.}
The online A/B testing results indicate the superiority and flexibility in real-world production environments. The code of our MLoRA is publicly available~\footnote{\href{https://github.com/gaohaining/MLoRA}{https://github.com/gaohaining/MLoRA}}.

\end{abstract}




\maketitle

\section{Introduction}
\begin{figure}
\includegraphics[width=\linewidth]{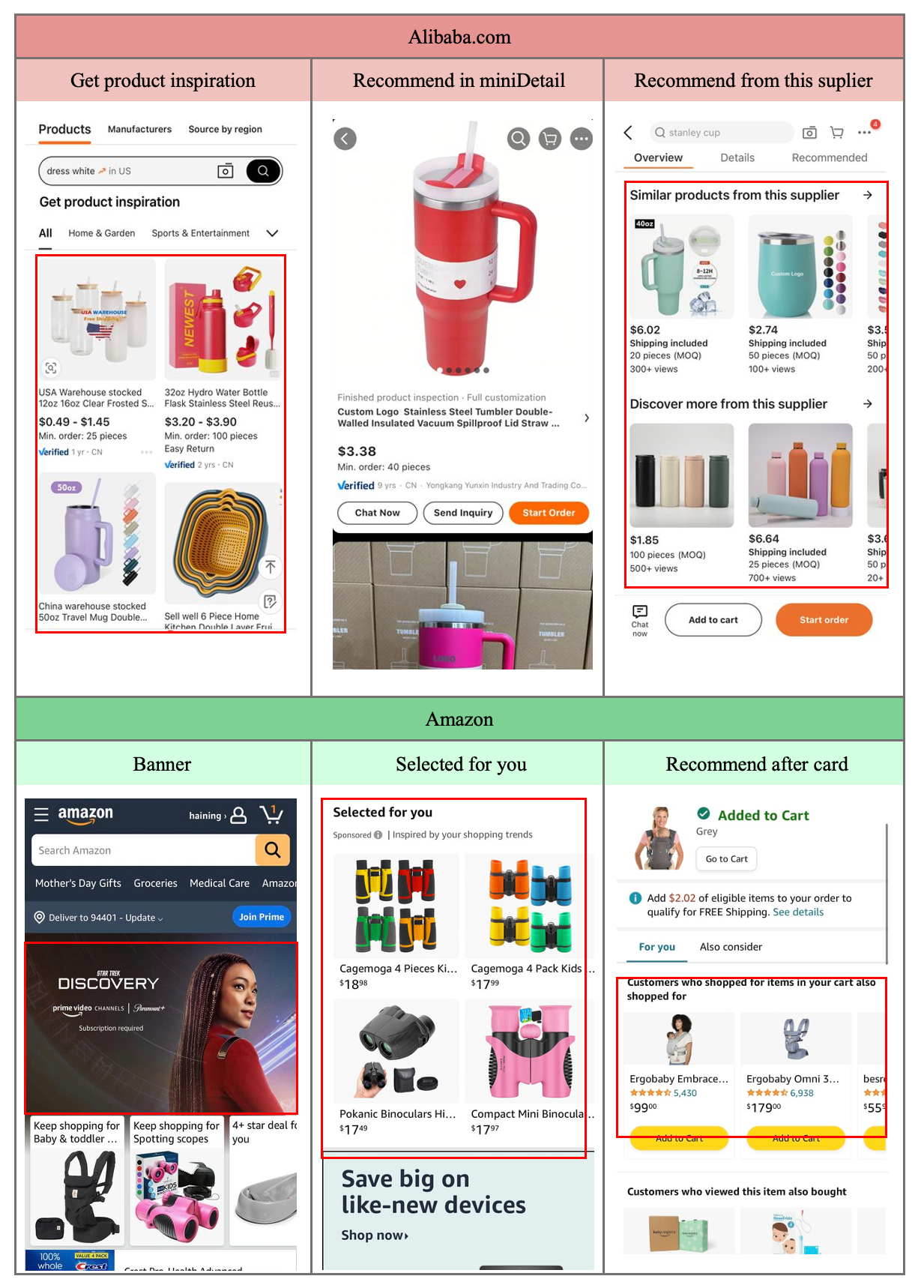}
\caption{Upper: three representative business domains of Alibaba.com, \textit{Get product inspiration}, \textit{Recommend in miniDetail} and \textit{Recommend from this supplier}. Lower: three key business sectors of Amazon, \textit{Banner}, \textit{Selected for you} and \textit{Recommend after card}.}
\label{fig:multidomainexp}
\end{figure}

Click-Through Rate (CTR) prediction is currently one of the most crucial estimation tasks, serving as a fundamental aspect in comprehending user behavior and driving personalized experiences~\cite{CTR_survey_IJCAI,PNN_ICDM,HOFM_NIPS}. 
It plays an essential role across various application areas such as personalized recommendation systems~\cite{Personalized_recommend_1,Personalized_recommend_2,Personalized_recommend_3,Personalized_recommend_4}, information retrieval~\cite{retrival_1,retrival_2,retrival_3,retrival_4}, and online advertising~\cite{CTR_survey_IJCAI,FinalMLP_AAAI,DIN_SIGKDD,advertise_1,advertise_2,advertise_3,advertise_4}. 
For instance, in the content recommendation field, platforms like TikTok leverage CTR prediction algorithms to curate personalized feeds for users, ensuring captivating and relevant browsing experiences~\cite{CTR_survey_IJCAI,FM_ICDM,WDL_DLRS,DeepFM_arxiv,Autoint_CIKM,PNN_ICDM}. 
While in the products recommendation area, platforms like Alibaba.com utilize CTR prediction to boost \ghn{conversion} rate which in turn increases website revenues~\cite{Xiao2020DMIN, Xiao2024DEI2N}.

For the sake of simplification, traditional CTR prediction approaches usually focus on one single domain, where the CTR model utilizes samples collected from a particular domain~\cite{WDL_DLRS, DeepFM_arxiv, Autoint_CIKM, NFM_SIGIR}.
However, the \textbf{multi-domain CTR prediction} is ubiquitous in reality, especially for large-scale platforms (e.g., Taobao or Amazon e-commerce platforms), which is still a hard nut to crack~\cite{multi-domain1,multi-domain2,multi-domain3}.
Taking Alibaba.com recommendation system as an example, there are hundreds of domains (e.g., \textit{Get product inspiration}, \textit{Recommend in miniDetail}, and \textit{Recommend from this supplier}) to satisfy the user demands and preferences. Each domain showcases a series of related products under a given topic as seen in the upper part of Figure ~\ref{fig:multidomainexp}. 
Another Amazon example can be seen in the lower part.

\begin{figure}[htbp]
\includegraphics[width=\linewidth]{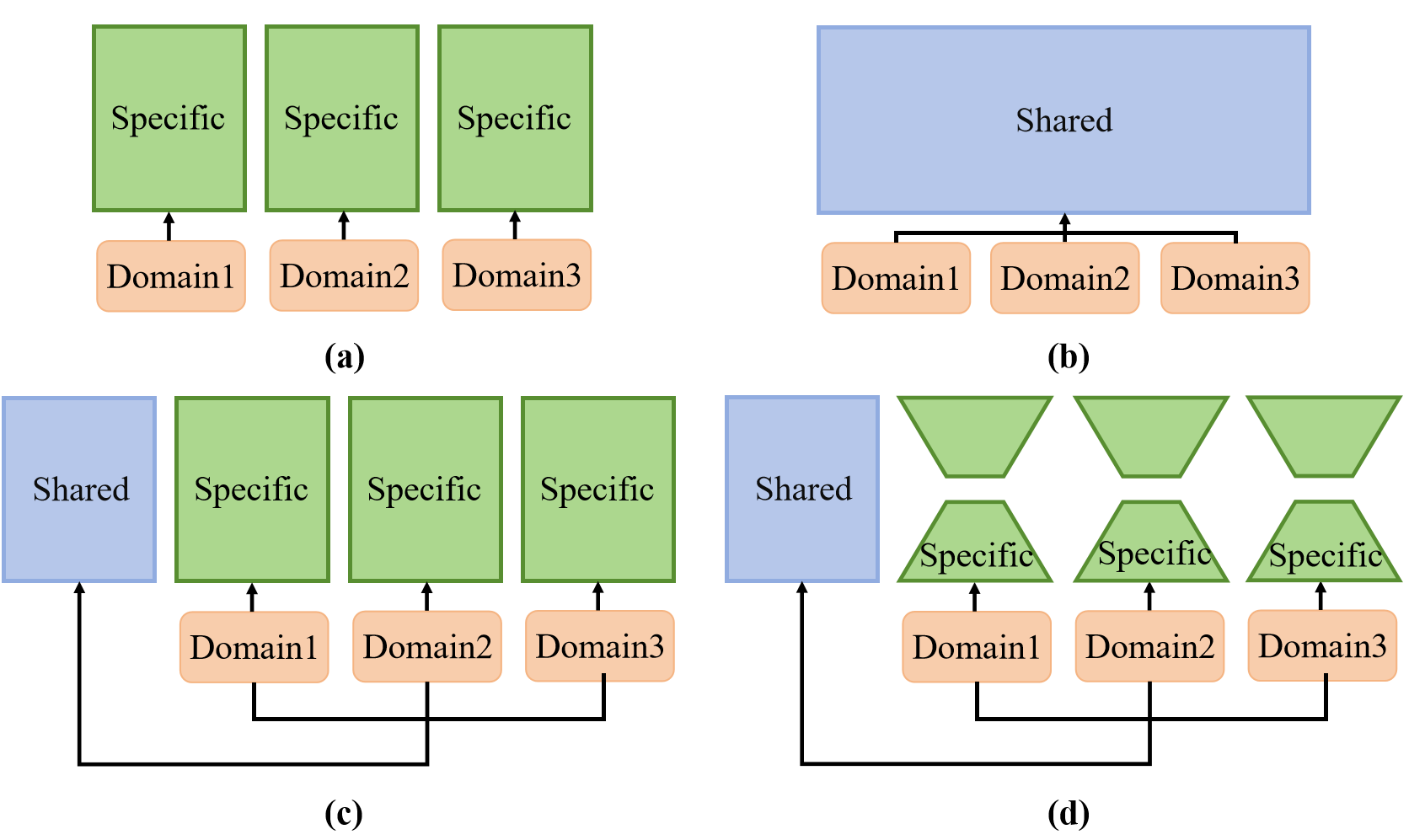} 
\caption{(a) Training separated model for each domain. (b) Training a unified model by mixing multi-domain data. (c) Multi-domain Framework~\cite{STAR_CIKM,MMoE_SIGKDD,PLE_recsys}. (d) Our Proposed MLoRA.}
\label{fig:ctr_model}
\end{figure}

Currently, two types of approaches are being researched to address the problem of multi-domain CTR prediction. 
The first type treats multiple domains independently and builds separate CTR prediction models for each domain as seen in Figure~\ref{fig:ctr_model} 
 (a)~\cite{WDL_DLRS,PNN_ICDM,DIN_SIGKDD}.
These approaches suffer from the problems of \textbf{data sparsity} since there is a limited volume of data for training a domain. 
Meanwhile, these approaches take less consideration of the relationships between multiple domains~\cite{WDL_DLRS, DeepFM_arxiv, NFM_SIGIR, FM_ICDM,Autoint_CIKM,PNN_ICDM}.
The second type trains a unified model by merging multi-domain data as seen in Figure~\ref{fig:ctr_model} (b)~\cite{Multi_domain_WITH_PARA_ML,HMoE_CIKM}. 
Although it solves the issue of data sparsity to a certain degree, this approach often fails to capture \textbf{domain diversity} as the mixed data neglects the differences in data distribution between domains.
In response to the above two issues of multi-domain recommendation, researchers have proposed corresponding solutions shown as Figure~\ref{fig:ctr_model} (c). 
Researchers~\cite{STAR_CIKM, MMoE_SIGKDD, PLE_recsys} alleviated the issue of sparse data by employing the shared part to learn an overall data distribution and introducing specific parts for each domain, thereby capturing unique data distribution information for each domain.
However, these approaches still lead to a sharp increase in \textbf{model parameters} and suffer from \textbf{insufficient learning} in some domains with small data volumes. 

Inspired by the success of Low-Rank Adaptor(LoRA) in large language model finetuning~\cite{LoRA_arxiv}, 
we propose the Multi-domain LoRA (i.e., MLoRA) approach for multi-domain CTR prediction, and its code is publicly available. 
The LoRA adaptor utilizes a low-rank structure to finetune Large Language Models (LLMs), effectively learning knowledge from domain data with fewer parameters.
Meanwhile, we build multiple LoRA adaptors for each domain, which learn data distribution of each domain more efficiently.  
Firstly, the Multi-domain LoRA approach effectively captures the unique data distributions of each domain, which facilitates the model in recognizing the diversity of information across different domains.
Secondly, the LoRA adaptor requires significantly fewer parameters, thus avoiding the potential issues of inadequate training on domains with sparse data and high computational cost.
The MLoRA approach is scalable and flexible to new domains by applying additional LoRA adaptors.
In addition, the proposed approach is a general framework, which is able to be applied to various CTR models such as WDL, PNN, NFM, etc~\cite{WDL_DLRS, NFM_SIGIR, Autoint_CIKM, PNN_ICDM, DCN_ADKDD, FiBiNET_RECSYS, DeepFM_arxiv, xDeepFM_SIGKDD}.

We have conducted extensive experiments on several public datasets. 
The experimental results indicate the superiority across multiple datasets by equipping current mainstream CTR models with MLoRA. 
\yzm{Furthermore, our MLoRA approach has been deployed on the Alibaba.COM e-commerce website, which integrated 10 core recommendation domains and the additional parameter size is only increased by $1.76\%$. }
It contributes to a 1.49\% increase in CTR and a 3.37\% increase in order conversion rate, as well as a 2.71\% \ghn{increase} in the number of paid buyers across the entire site.
The main contributions of this paper are summarized as follows:
\begin{itemize}
    \item We propose a novel approach MLoRA to address the problem of multi-domain CTR prediction by incorporating LoRA adaptors. 
    The MLoRA approach is model-agnostic, which can be applied to nearly all deep learning-based CTR approaches.
    \item We introduce multiple LoRA networks, where the integration of mixed data mitigates data sparsity, while the implementation of multiple LoRA adaptors captures the diversity within each domain.
    \item We evaluate our proposed MLoRA on several datasets with state-of-the-art methods. Extensive experimental results demonstrate the effectiveness and superiority of our MLoRA approach.
    \item We deploy MLoRA in a real-world production e-commerce website. The A/B testing results verify its advancement in CTR and conversion rate only at a small cost of parameter increase.
\end{itemize}

\section{Related Work}
In this section, we will review the related work from the following two respects: CTR Prediction and Low-Rank Adaptor.

\subsection{CTR Prediction}
The prediction of Click-Through Rate is a fundamental task in recommendation, advertising and search systems, where an effective CTR prediction model is crucial for enhancing commercial value.
We will describe existing methods in two groups: Single-Domain Approaches and Multi-Domain Approaches.

\textbf{Single-Domain Approaches.}
In recent years, the methodology for CTR tasks has shifted from traditional shallow~\cite{t1,t2,t3,FM_ICDM,t5} approaches to deep learning techniques~\cite{CTR_survey_IJCAI, WDL_DLRS, DeepFM_arxiv, PNN_ICDM, DIN_SIGKDD, dien}. 
Deep learning-based CTR models typically follow an architecture consisting of embedding and multi-layer perceptron (MLP) layers.
The embedding layer initially transforms sparse ID features into dense vectors and continuous numeric features are also discretized through bucketing before embedding. 
The MLP layer takes all embeddings as inputs to perform deep feature interactions.
PNN~\cite{PNN_ICDM} adds a product layer between the embedding and MLP layers to explicitly carry out feature interactions.
WDL~\cite{WDL_DLRS} combines a linear layer for memorizing information on the wide side with a deep MLP network for generalization on the deep side.
DCN~\cite{DCN_ADKDD} introduces a cross-network to facilitate feature interactions on the wide side, while DeepFM~\cite{DeepFM_arxiv} incorporates a Factorization Machine to boost the capability for feature interaction on the wide side.
NFM~\cite{NFM_SIGIR} improves the deep side by introducing the Bi-Interaction Layer structure to process second-order crossing information, enabling deep MLP networks to better capture high-order combinations between features.
Another innovation for the multi-layer MLP is AFM~\cite{AFM}, which implements an attention mechanism, allowing each feature interaction to be weighted differently, thereby filtering valuable combination information.
AutoInt~\cite{Autoint_CIKM} and CAN~\cite{CAN} further propose the self-attention mechanism for comprehensive feature interactions.
FiBiNET~\cite{FiBiNET_RECSYS} dynamically learns feature importance and employs a Bilinear method to learn fine-grained feature interactions. 
xDeepFM~\cite{xDeepFM_SIGKDD} builds on the WDL~\cite{WDL_DLRS} model by introducing the Compressed Interaction Network (CIN) module, which generates vector-wise level feature interactions explicitly. 
The aforementioned methods are mainly from the perspective of feature interactions, another modeling approaches start from the angle of user behavior modeling. 
DIN~\cite{DIN_SIGKDD} innovatively incorporates the user behavior sequence and employs an attention mechanism to capture diverse interests related to the target item. 
DIEN~\cite{dien} introduces GRU to model the evolution and transformation of user interests over time. 
MIMN~\cite{MIMN} proposes a novel memory-based architecture, which is designed to capture user interests from long sequential behavior sequences. 
It provides the capability to process sequences extending to thousands in length.

 \textbf{Multi-Domain Approaches.}
In real-world applications, data is often collected from multiple domains. 
Multi-domain data typically presents two challenges: firstly, some domain data is sparse, which cannot meet the requirements for model training. 
Secondly, there are differences in data between different domains, making it difficult to apply a single-domain approach rigidly.
Facing the challenge of CTR prediction in multi-domain data scenarios, researchers have proposed some solutions. 
Domain generalization (DG) methods~\cite{DG1} offer a solution approach to the multi-domain CTR prediction problem by extracting common knowledge from multiple domains and learning features that generalize to unknown domains. 
Li et al.~\cite{DG2} suggest that this approach can effectively alleviate the issue of data sparsity.
STAR~\cite{STAR_CIKM} proposed Partitioned Normalization(PN) and domain-specific FCN to capture the uniqueness of each domain's features.
MMOE~\cite{MMoE_SIGKDD} adopts a Mixture of Experts (MoE) architecture, sharing expert modules across all domains while training a gating network for each domain. 
PLE~\cite{PLE_recsys} achieves joint learning of information from different domains by explicitly separating shared and task-specific components.
However, the task-specific component is almost the same as the shared part, it will cause a significant increase in model parameters. 
Moreover, it still suffers from insufficient training in sparse data domains.

\subsection{Low-Rank Adaptor}
Over the past year, the pretraining\cite{pretrain_1,pretrain_2,pretrain_3,pretrain_4} and finetuning\cite{fintune_1, fintune_2, fintune_3,fintune_4, fintune_5} architecture has become a fundamental modeling approach in the NLP field.
Pretraining allows the model to acquire generalizable knowledge from a large and diverse dataset, while finetuning makes the model more closely match the data distribution of a specific domain. 
However, in many domains, the amount of data available for finetuning is very limited, sometimes even much less than the number of parameters in the model, making it a challenge for finetuning tasks to learn sufficiently. 
In light of this, a new perspective based on analyzing intrinsic dimensionality~\cite{intrinsic_dim} is proposed. The intrinsic dimension describes the minimum number of dimensions needed to solve the problem.
Measuring the intrinsic dimension will tell us how many free parameters are required to closely approximate the solution to the problem while finetuning for each target task.
Studies~\cite{intrinsic_dim,low_rank} show that in some cases, the intrinsic dimension is several orders of magnitude smaller than the full parameterization. 

Therefore, a finetuning approach named LoRA~\cite{LoRA_arxiv} is proposed, which freezes the pretrained model weights and injects trainable rank decomposition matrices, named $\mathbf{A}$ and $\mathbf{B}$, into each layer.
Benefiting from the low-rank, the number of parameters in the LoRA module that need to be finetuned is significantly lower than that of the pretrained parameters.
This enables downstream tasks to be finetuned very efficiently, achieving commendable results while also reducing the amount of training data required.
A pretrained model can be shared and employed to construct numerous small LoRA modules for various tasks across different domains. 
By freezing the shared model and efficiently switching tasks through the replacement of low-rank matrices $\mathbf{A}$ and $\mathbf{B}$, we can significantly reduce the storage requirements and the overhead related to task switching.
\begin{figure*}[htbp]
\includegraphics[width=\linewidth]{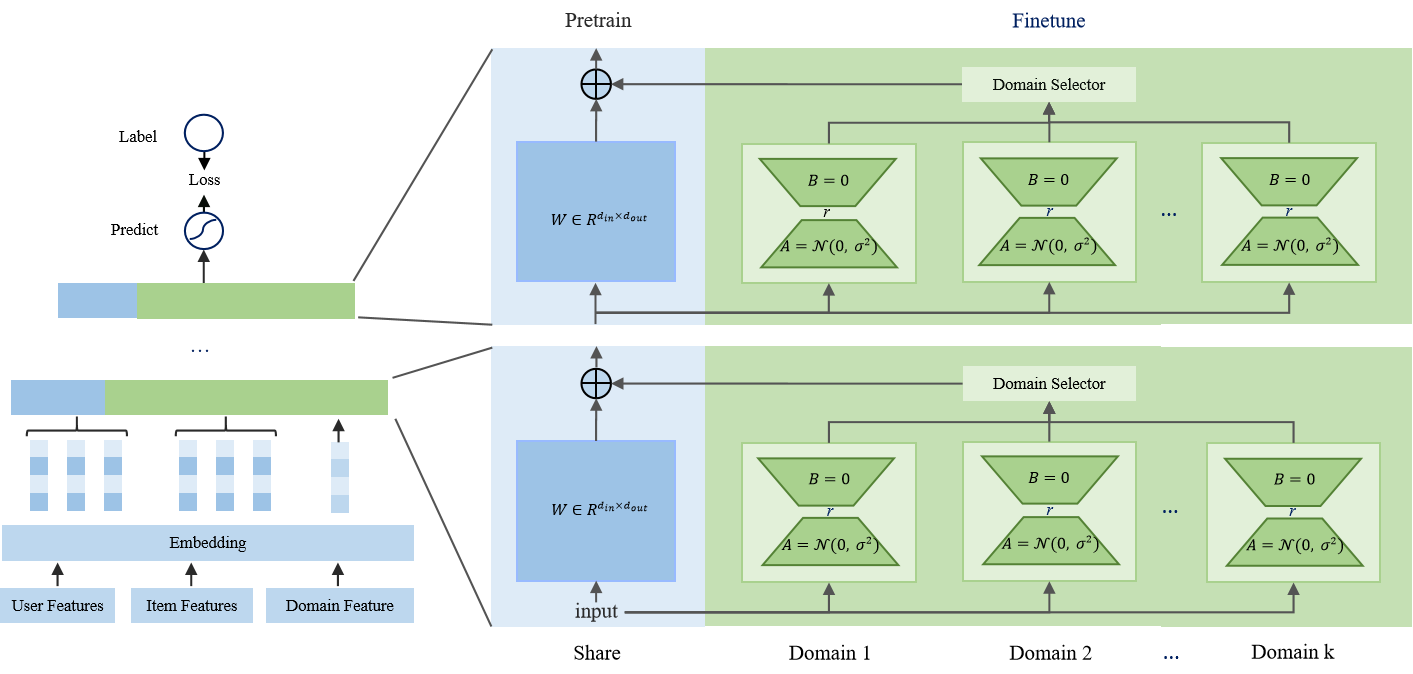}
\caption{The overview architecture of the proposed MLoRA, which consists of input features,  embedding layer, shared network, multi-domain network, and domain selector. 
Each domain network consists of low-rank matrix $\mathbf{A}$ and $\mathbf{B}$, and the outputs of the domain network and share network are added together to serve as the output of the entire layer. The embedding layer and shared network are trained during the pretraining phase, while the domain network is trained during the finetuning phase.
}
\label{fig:MLoRA}
\end{figure*}

\section{Methodology}
In this section, we present the proposed MLoRA in detail. 
Figure~\ref{fig:MLoRA} illustrates the overall network architecture of our model.

\subsection{Problem definition}
We define the notations and problem settings in this paper. The model uses sparse/dense inputs such as user features, item features, context features, and so on. 
The predicted target  $\hat{y}$ is the probability that user $u$ will interact with item $i$, which is calculated via:
\begin{equation}\label{eqn-1} 
\hat{y} = F(E(u_1), ..., E(u_n); E(i_1), ..., E(i_m); E(c_1), ..., E(c_k)),
\end{equation}
where $\{u_1, ..., u_n\}$  is the set of user features. 
$\{i_1, ..., i_m\}$ is the set of target item feature and $\{c_1, ..., c_k\}$ is the set of context features. 
The $E(i) \in R^d$ means the embedding layer, which maps the sparse feature into learnable dense vectors. 
Continuous numeric features are discretized through bucketing before embedding.

When jointly modeling across multiple domains, we need to integrate the data from all domains and take the domain ID as input, which is calculated via:
\begin{equation}\label{eqn-2} 
\hat{y_t} = F(E(u_1), ..., E(u_n); E(i_1), ..., E(i_m); E(c_1), ..., E(c_k); E(t)),
\end{equation}
where $\hat{y_t}$  is the probability that user $u$ will interact with item $i$ in domain $t$ (i.e., the domain ID). 

\subsection{Low-Rank Adaptor}
LoRA is a finetuning approach based on an improvement of a single fully connected layer which has a full-rank parameter matrix $\mathbf{W} \in R^{d_{out}{\times}d_{in}}$,  where $d_{in}$ is the dimension of the input and $d_{out}$ is the dimension of the output. 
Assuming that after finetuning, the parameter matrix $\mathbf{W}$ is updated to $\mathbf{W'}$. 
Without loss of generality, finetuning can be described as learning $\Delta{\mathbf{W}}$, where
\begin{equation}\label{eqn-3} 
\Delta{\mathbf{W}} = \mathbf{W}^\prime - \mathbf{W},
\end{equation}
When finetuning a large pretrained model for a specific task, the full-rank matrix often has considerable parameter redundancy\cite{intrinsic_dim,low_rank,LoRA_arxiv}, with a rank $r \ll min(d_{in}, d_{out})$, then $\Delta{\mathbf{W}}$ can be represented as :
\begin{equation}\label{eqn-4} 
\Delta{\mathbf{W}} = \mathbf{W}^\prime - \mathbf{W} = \mathbf{B}\mathbf{A},
\end{equation}
where $\mathbf{B} {\in} R^{d_{out} {\times} r}$ and $\mathbf{A} {\in} R^{r {\times} d_{in}}$.  
It is particularly important to note that in order to ensure that $\Delta{\mathbf{W}}$ is zero at the beginning of training, $\mathbf{A}$ is initialized with a Gaussian distribution, and $\mathbf{B}$ is initialized with zeros.
Therefore, the forward pass of LoRA can be represented as:
\begin{equation}\label{eqn-5} 
\mathbf{h} = \mathbf{W}^\prime \mathbf{x} = \mathbf{W}\mathbf{x} + \mathbf{B}\mathbf{A}\mathbf{x},
\end{equation}
where $\mathbf{x}$ is the input and $\mathbf{h}$ is the output.

\subsection{MLoRA for Multi-domain CTR}
When training a CTR model, there are two parts that need to be fitted: the generalizable common information and the personalized distinctive information. 
Thus, the predicted target  $\hat{y}$  can be calculated via:
\begin{equation}\label{eqn-6} 
\hat{y} = F(\textbf{x}) = F_0(\textbf{x}) + \Delta{F}(\textbf{x}),
\end{equation}
Considering the sparsity of individual domains, a model $L(x)$ utilizing low-rank matrices as parameters can be used to learn personalized information for each domain. 
Then the $\hat{y}$ can be calculated via:
\begin{equation}\label{eqn-7} 
\hat{y} = F_0(\textbf{x}) + L(\textbf{x}),
\end{equation}
When modeling across different domains, the primary focus is on the commonalities and differences between domains. 
Therefore, a low-rank matrix model is constructed separately for each domain. 
Then the $\hat{y_t}$ can be calculated via:
\begin{equation}\label{eqn-8} 
\hat{y_t} = F_0(\textbf{x}) + L_t(\textbf{x}),
\end{equation}
From this perspective, in order to achieve a more comprehensive fitting, we propose the approach named MLoRA, as shown in Figure ~\ref{fig:MLoRA}, which splits each layer of the model into a common part and a personalized part, rather than treating the model as a whole. 
A single layer can be represented as:
\begin{equation}\label{eqn-9} 
\textbf{h}_t = \textbf{W}\textbf{x} + \Delta\textbf{W}_t = \mathbf{W}\textbf{x} + \textbf{B}_t \textbf{A}_t \textbf{x},
\end{equation}
Unlike NLP models, CTR models usually exhibit significant differences in network width between layers, hence using a fixed $r$ value across different layers is not equitable. 
In the MLoRA approach, a fixed scaling factor, referred to as a temperature coefficient $\alpha$, is employed. 
Thus $r$ can be calculated via:
\begin{equation}\label{eqn-10} 
r = \max(\frac{d_{out}}{\alpha},  1).
\end{equation}

MLoRA applies a two-phase training strategy. 
In the pretraining phase, the backbone network is trained with large-scale pretraining data to learn information that can be generalized across various domains. 
During the finetuning phase, we add the MLoRA networks while freezing the backbone network. 
The finetuning phase is focused solely on $\textbf{A}$ and $\textbf{B}$ updates to learn personalized information for each domain.

\section{Experiments}
Extensive experiments are conducted to evaluate our MLoRA approach.
The datasets, baselines, and experimental settings are described in Section~\ref{sec::datasets}, Section~\ref{sec::baselines}, and Section~\ref{sec::setting}, respectively.
In Section ~\ref{sec::mainexp}, we detail the experimental design for evaluating MLoRA, conducting thorough experiments across three datasets and performing detailed analyses of the results. 
Finally, in Section~\ref{Sec:Industry_App}, we discuss real-world production applications and online A/B testing results.

\subsection{Datasets}\label{sec::datasets}
In our experiments, we construct multi-domain recommendation benchmark datasets based on three publicly available real-world datasets: Taobao~\cite{Taobao-dataset}, Amazon~\cite{Amazon-dataset}, and Movielens~\cite{Amazon-dataset}. The statistics of them are summarized in Table~\ref{table:dataset}.

The Taobao dataset consists of user click logs from various themed scenarios in Taobao APP~\footnote{https://world.taobao.com/}. 
These scenarios are already partitioned into domains based on different thematic topics, such as \textit{what to take when traveling}, \textit{how to dress up yourself for a party} and \textit{things to prepare when a baby is coming}, etc. We randomly select 10, 20, and 30 domains to construct multi-domain datasets which are named \textbf{Taobao-10}, \textbf{Taobao-20}, and \textbf{Taobao-30}.

The Amazon dataset is a large-scale dataset comprising reviews (ratings, text, and helpfulness votes) collected from Amazon.com~\footnote{https://www.amazon.com/}, along with product metadata (descriptions, category information, price, brand, and image features), and links (also viewed/also bought graphs). 
We split domains based on product categories such as Musical Instruments and Video Games. 
We select 6 domains to form the dataset \textbf{Amazon-6}.

The Movielens dataset, published in February 2003, contains 1 million movie ratings collected from 6000 users across 4000 movies. 
We partition the Movielens dataset using gender as a categorical feature, forming the multi-domain dataset \textbf{Movielens-gen}.

\begin{table*}
\centering
\caption{Overall statistics of the datasets.}
\label{table:dataset}
\begin{tabular*}{0.6\linewidth}{@{\extracolsep{\fill}}ccccccc}
\toprule
\textbf{Dataset} & Domain & User & Item & Train & Validation & Test\\
\midrule
Amazon-6 & 6 & 445789 & 172653 & 9968333 & 3372666 & 3585877\\
\midrule
Taobao-10 & 10 & 23778 & 6932 & 92137 & 37645 & 43502\\
Taobao-20 & 20 & 58190 & 16319 & 243592 & 96591 & 106500\\
Taobao-30 & 30 & 99143 & 29945 & 394805 & 151369 & 179252\\
\midrule
Movielens-gen & 2 & 6041 & 3953 & 600125 & 200042 & 200042\\
\bottomrule
\end{tabular*}
\end{table*}

\begin{table*}
\centering
\caption{The WAUC ($\%$) results of CTR prediction on different datasets. Note Base refers to the original results of the corresponding methods and Base+MLoRA refers to the results
with the assistance of MLoRA. Avg is the average results across all approaches. $\Delta$ refers to the improvement of
Base+MLoRA compared to Base.}
\label{table:main_exp_result}
\begin{tabular*}{\linewidth}{@{}c|c|c|c|c|c|c|c|c|c|c|c|c@{}}
\toprule
\textbf{Dataset} & \textbf{Approach} & MLP & STAR & WDL & NFM & AutoInt & PNN & DCN & FiBiNET & DeepFM & xDeepFM & Avg\\
\midrule
\multirow{3}{*}{\bf Taobao-10}
                         & Base & 72.92 & 75.51 & 73.12 & 76.88 & 75.56 & 76.42 & 72.07 & 76.51 & 75.01 & 75.26 & 74.93\\
                         & Base+MLoRA & \bf 74.53 & \bf 76.17 & \bf 73.51 & \bf 77.08 & \bf 75.83 & \bf 76.74 & \bf 72.14 & \bf 76.77 & \bf 75.66 & \bf 75.64 & \bf 75.41\\
                         & $\Delta$ & +1.61 & +0.66 & +0.39 & +0.20 & +0.27 & +0.32 & +0.07 & +0.26 & +0.65 & +0.38 & +0.48\\
                        \midrule
\multirow{3}{*}{\bf Amazon-6}
                         & Base & 75.07 & 77.26 & 73.25 & 65.07 & 74.10 & 74.03 & 75.61 & 74.93 & 73.55 & 74.64 & 73.75\\
                         & Base+MLoRA & \bf 77.48 & \bf 77.28 & \bf 74.25 & \bf 67.41 & \bf 74.46 & \bf 75.18 & \bf 75.62 & \bf 75.27 & \bf 74.08 & \bf 74.83 & \bf 74.58\\
                         & $\Delta$ & +2.41 & +0.02 & +1.00 & +2.34 & +0.36 & +1.15 & +0.01 & +0.34 & +0.53 & +0.19 & +0.83\\
                        \midrule
\multirow{3}{*}{\bf Movielens-gen}
                         & Base & 80.15 & 80.18 & 80.17 & 80.57 & 80.25 & 80.26 & 80.28 & 80.47 & 80.24 & 80.26 & 80.28\\
                         & Base+MLoRA & \bf 80.39 & \bf 80.36 & \bf 80.41 & \bf 80.59 & \bf 80.41 & \bf 80.39 & \bf 80.42 & \bf 80.74 & \bf 80.46 & \bf 80.44 & \bf 80.46\\
                         & $\Delta$ & +0.24 & +0.18 & +0.24 & +0.02 & +0.16 & +0.13 & +0.14 & +0.27 & +0.22 & +0.18 & +0.18 \\
                         \midrule
            {\bf - } & Avg($\Delta$) & +1.42 & +0.29 & +0.54 & +0.85 & +0.26 & +0.53 & +0.07 & +0.29 & +0.47 & +0.25 & +0.50 \\
\bottomrule
\end{tabular*}
\end{table*}

\subsection{Baselines}\label{sec::baselines}
Both state-of-the-art single-domain and multi-domain methods are selected as baselines.
For fairness, the hidden layers in all models are set to [256, 128, 64].
\begin{itemize}
\item \textbf{MLP} is a basic neural network model, consisting of multiple fully connected layers linked sequentially.
\item \textbf{WDL}~\cite{WDL_DLRS} combines traditional feature engineering with deep models to enable the model to possess characteristics of both memorization and generalization.
\item \textbf{NFM}~\cite{NFM_SIGIR} utilizes the Bi-Interaction Layer structure to process second-order crossing information, allowing the DNN structure to better learn the information of cross features and reduce the difficulty of learning higher-order cross-feature information.
\item \textbf{AutoInt}~\cite{Autoint_CIKM} achieves automatic feature cross through ingenious model design by leveraging multi-layered multi-head self-attention modules stacked together combined with residual network to extract low-order and high-order cross-information from input features.
\item \textbf{PNN}~\cite{PNN_ICDM} uses embedding layers to learn distributed representations of the data, captures interactive patterns between inter-field categories using product layers, and ultimately explores high-dimensional feature interactions using fully connected layers.
\item \textbf{DCN}~\cite{DCN_ADKDD} proposes a cross-network to explicitly encode features for better model representation.
\item \textbf{FiBiNET}~\cite{FiBiNET_RECSYS} dynamically learns the importance of features using the squeeze-excitation network structure and models cross features better using a bilinear function.
\item \textbf{DeepFM}~\cite{DeepFM_arxiv} integrates the advantages of factorization machines and deep neural networks by simultaneously considering interactions between lower-order and higher-order features, enabling end-to-end learning of all feature interactions.
\item \textbf{xDeepFM}~\cite{xDeepFM_SIGKDD} designs a compressed interaction network to explicitly learn higher-order feature interactions, combining it with traditional DNNs to jointly learn explicit and implicit higher-order features.
\item \textbf{STAR}~\cite{STAR_CIKM} is a multi-domain recommendation method that divides parameters into shared and specific parts. Meanwhile, it proposes Partitioned Normalization to integrate differentiated information across different domains.
\end{itemize}

\subsection{Settings}\label{sec::setting}
For a fair comparison, we implement all CTR prediction approaches on TensorFlow 1.12 and run them on two RTX 2080Ti GPUs. 
For the Taobao dataset, as the input embeddings are provided already, there is no need to train the input embedding.
For Amazon and Movielens datasets, the input embedding size is set as 8.
We set the learning rate to 0.001, dropout rate to 0.5, batch size to 1024, and employed binary cross-entropy~\cite{cross_entropy} as the loss function.
All approaches employ a three-layer feedforward neural network with hidden layer sizes of [256, 128, 64], and temperature coefficient $\alpha$ is set to 32.
\gdh{We divided the dataset into training, validation, and testing sets, where about three-fifths of the dataset is randomly selected as the training set, and about one-fifth is randomly chosen as the validation set, and the rest as the testing set following~\cite{MAMDR}.}



Our training strategy will save the parameters of the model with the best performance during training. 
Training would stop when the model's performance failed to exceed the best performance achieved in previous training epochs for a consecutive number of times. 
It is determined by a patience parameter following ~\cite{MAMDR}.
In practice, most training schemes set patience to $3$. 
Due to the possibility of some schemes terminating training prematurely, to ensure fairness, we set patience=$5$ to maintain consistency across all schemes. 

AUC (Area Under the ROC Curve) is one of the most commonly used metrics for evaluating CTR prediction~\cite{AUC}.
Note 0.1\% absolute AUC gain is regarded as significant for the CTR task~\cite{DIN_SIGKDD,WDL_DLRS,Autoint_CIKM}.
As the data scale varies across different domains in the dataset, we utilized WAUC (Weighted AUC)~\cite{WAUC} as the metric for evaluating model performance:
\begin{equation}\label{eqn-10} 
WAUC=\sum_{i=1}^{n}{\omega_i * AUC_i },
\end{equation}
where $AUC_i$ refers to the AUC result for the $i$-th domain and $\omega_i$ represents the proportion of the data volume for the $i$-th domain compared to the total data volume. 
It is defined as
\begin{equation}\label{eqn-12} 
\omega_i=\frac{s_i}{\sum_{i=1}^{n}{s_i}} \times 100 \%,
\end{equation}
where $s_i$ represents the size of the data for the $i$-th domain.

\subsection{Main Experiments}\label{sec::mainexp}



We conduct extensive experiments on three public datasets: Taobao-10, Amazon-6, and Movielens-gen as shown in Table~\ref{table:dataset}.
The MLoRA approach is model-agnostic, which allows it to be applied to various CTR prediction methods. 
Therefore, to evaluate the performance of MLoRA, we applied it to a wide range of existing deep CTR prediction models based on deep learning and compared the results of the original models (denoted as Base) with those equipped with MLoRA. 
Here, WAUC(\%) score is reported as the metric.

\textbf{Observation 1}: From Table.~\ref{table:main_exp_result}, it can be observed that with the assistance of MLoRA, all approaches achieve performance improvements on the Taobao-10, Amazon-6, and Movielens-gen datasets. 
The improvements across all methods on the three datasets are 0.49\%, 0.83\%, and 0.18\%, respectively, with an average of 0.5\%.
These improvements are statistical significant by applying a paired T-test.
This may be due to the base models (except for STAR) training a single network by mixing multi-domain data, which fails to capture the diversity among domains. 
Introducing LoRA adaptors helps alleviate this limitation. 
The STAR approach copies the shared network to each specific domain and trains the domain network again. 
Due to the large parameter size of the domain network and data sparsity, the overall model tends to be under-fitting.
Therefore, introducing MLoRA can also improve the performance of STAR by reducing the complexity and parameterization of specific FCNs through the introduction of LoRA.
This claims that MLoRA is a model-agnostic framework that can be easily deployed on various deep CTR prediction networks. 
It effectively alleviates data sparsity issues and captures inter-domain diversity information, demonstrating good transferability and feasibility.

\textbf{Observation 2}: By comparing the results of MLoRA on three datasets, we find that on the Amazon-6 dataset, MLoRA achieves the highest average performance improvement of 0.83\% (from 73.75\% to 77.48\%). 
We attribute this to the larger dataset size of Amazon-6 and the more pronounced differences between domains, which the base model fails to learn sufficiently.
Introducing specific LoRA adaptors for each domain allows the model to capture differentiated information better, leading to a significant performance boost.
Furthermore, the performance improvement on the Movielens dataset is the lowest at 0.18\% (from 80.28\% to 80.46\%), with all base models achieving similar performance levels, ranging from 80.1\% to 80.6\% on Movielens-gen. 
Even after deploying MLoRA, the performance remains in the range of 80.3\% to 80.8\%. 
This is because the Movielens dataset has a smaller data size, and the complexity and parameterization of the base models are already sufficient for the CTR prediction task on this dataset. 
Additionally, the differences between domains are relatively small, resulting in minimal improvement with MLoRA.

\begin{table*}[ht]
\centering
\caption{On Taobao-10 dataset, we obtained the AUC ($\%$) results for each domain using the MLP, FiBiNET, and DeepFM schemes. }
\label{table:domain}
\begin{tabular*}{\linewidth}{c|p{1cm}<{\centering}|p{1cm}<{\centering}|p{1cm}<{\centering}|p{1cm}<{\centering}|p{1cm}<{\centering}|p{1cm}<{\centering}|p{1cm}<{\centering}|p{1cm}<{\centering}|p{1cm}<{\centering}|p{1cm}<{\centering}|c}
\toprule
\textbf{Approach} & \textbf{0} & \textbf{1} & \textbf{2} & \textbf{3} & \textbf{4} & \textbf{5} & \textbf{6} & \textbf{7} & \textbf{8} & \textbf{9} & \textbf{WAUC}\\
\midrule

\textbf{MLP} & 69.09 & 58.23 & 68.17 & 77.29 & 79.70 & 74.21 & 56.56 & 75.47 & 64.43 & 70.21 & 72.92 \\
\textbf{MLP+MLoRA} & \bf 70.40 & \bf 62.24 & \bf 69.03 & \bf 79.11 & \bf 80.87 & \bf 74.47 & \bf 60.12 & \bf 76.61 & \bf 65.90 & \bf 72.46 & \bf 74.53 \\
\textbf{$\Delta$} & +1.31 & +4.02 & +0.86 & +1.82 & +1.17 & +0.26 & +3.55 & +1.14 & +1.47 & +2.25 & +1.61 \\

\midrule

\textbf{FiBiNET} & 71.90 & 63.13 & 70.43 & 79.28 & 81.19 & 78.11 & 68.04 & 82.54 & 65.42 & 76.08 & 76.51 \\
\textbf{FiBiNET+MLoRA}& \bf 72.56 & \bf 63.70 & \bf 70.65 & \bf 79.35 & \bf 81.20 & \bf 78.58 & \bf 68.58 & \bf 83.25 & \bf 65.55 &  \bf 76.08 & \bf 76.77  \\
\textbf{$\Delta$}& +0.67 & +0.57 & +0.22 & +0.07 & +0.01 & +0.47 & +0.54 & +0.71 & +0.13 & +0.00 & +0.26 \\
\midrule

\textbf{DeepFM} & 72.44 & 64.27 & 66.60 & 76.46 & 81.36 & 76.56 & 66.85 & 83.53 & 67.34 & 72.69 & 75.01 \\
\textbf{DeepFM+MLoRA} & \bf 72.81 & \bf 64.41 & \bf 67.21 & \bf 77.80 & \bf 81.65 & \bf 77.41 & \bf 67.11 & \bf 83.65 & \bf 67.58 & \bf 72.79 & \bf 75.66 \\
\textbf{$\Delta$}& +0.37 & +0.14 & +0.61 & +1.34 & +0.29 & +0.85 & +0.26 & +0.12 & +0.24 & +0.10 & +0.65 \\

\midrule

 Avg($\Delta$) & +0.78 & +1.57 & +0.56 & +1.08 & +0.49 & +0.53 & +1.45 & +0.66 & +0.61 & +0.78 & +0.84\\
\bottomrule
\end{tabular*}
\end{table*}

\textbf{Observation 3}: To demonstrate the effectiveness of MLoRA's LoRA adaptors on each domain, we analyze the performance improvement of deep models on each domain of Taobao-10 dataset.
As MLP serves as the most basic model structure, and FiBiNET and DeepFM exhibit relatively lower and higher performance improvements on \ghn{the} Taobao-10 dataset, respectively, we select these three approaches for analysis. The results are presented in Table~\ref{table:domain}. 

It can be observed that MLoRA enhances \ghn{consistent} improvement across all domains, with the most significant improvements observed in Domain\_1 and Domain\_6. 
One of the reasons is that Domain\_1 and Domain\_6 have smaller data sizes and greater dissimilarities compared with other domains.
Therefore, we conclude that by introducing LoRA adaptors, we effectively alleviate the issues of data sparsity and domain diversity. 
Particularly, the performance improvement becomes more pronounced when these issues are severe. 
The same phenomenon is observed for \ghn{the} Amazon-6 and \ghn{the} Movielens-gen. Thus, it can be concluded that when the dataset size increases and the distributional differences between domains become more pronounced, MLoRA achieves better results. 
This highlights its ability to not only learn overall data distribution information but also capture relational information between different domains.

\begin{table*}
\centering
\caption{The WAUC ($\%$) results of CTR prediction on a multi-domain Taobao dataset classified by 10, 20, and 30 domains.}
\label{table:Ablation-taobao}
\begin{tabular*}{\linewidth}{@{}c|c|c|c|c|c|c|c|c|c|c|c|c@{}}
\toprule
\textbf{Dataset} & \textbf{Approach} & MLP & STAR & WDL & NFM & AutoInt & PNN & DCN & FiBiNET & DeepFM & xDeepFM & Avg\\
\midrule
\multirow{3}{*}{\bf Taobao-10}
                         & Base & 72.92 & 75.51 & 73.12 & 76.88 & 75.56 & 76.42 & 72.07 & 76.51 & 75.01 & 75.26 & 74.93\\
                         & Base+MLoRA & \bf 74.53 & \bf 76.17 & \bf 73.51 & \bf 77.08 & \bf 75.83 & \bf 76.74 & \bf 72.14 & \bf 76.77 & \bf 75.66 & \bf 75.64 & \bf 75.41\\
                         & $\Delta$ & +1.61 & +0.66 & +0.39 & +0.20 & +0.27 & +0.32 & +0.07 & +0.26 & +0.65 & +0.38 & +0.48\\
                        \midrule
\multirow{3}{*}{\bf Taobao-20}
                         & Base & 77.05 & 77.89 & 77.33 & 78.55 & 77.50 & 79.13 & 79.32 & 79.27 & 79.52 & 79.13 & 78.27\\
                         & Base+MLoRA & \bf 77.54 & \bf 78.66 & \bf 77.83 & \bf 78.84 & \bf 78.09 & \bf 79.21 & \bf 79.49 & \bf 79.29 & \bf 79.81 & \bf 79.45 & \bf 78.62\\
                         & $\Delta$ & +0.49 & +0.77 & +0.50 & +0.29 & +0.59 & +0.08 & +0.17 & +0.02 & +0.29 & +0.32 & +0.35\\
                        \midrule
\multirow{3}{*}{\bf Taobao-30}
                         & Base & 76.96 & 77.03 & 77.74 & 78.09 & 78.34 & 78.66 & 75.81 & 76.38 & 76.13 & 79.27 & 77.44\\
                         & Base+MLoRA & \bf 77.15 & \bf 78.29 & \bf 77.90 & \bf 78.32 & \bf 78.36 & \bf 78.72 & \bf 76.46 & \bf 76.59 & \bf 76.25 & \bf 79.29 & \bf 77.73\\
                         & $\Delta$ & +0.19 & +1.26 & +0.16 & +0.23 & +0.02 & +0.06 & +0.65 & +0.21 & +0.12 & +0.02 & +0.29 \\
                         \midrule
 {\bf - } & Avg($\Delta$) & +0.76 & +0.90 & +0.35 & +0.24 & +0.29 & +0.15 & +0.30 & +0.16 & +0.35 & +0.24 & +0.37\\
\bottomrule
\end{tabular*}
\end{table*}

\subsection{Ablation Studies}\label{sec::ablation}
\begin{figure}[ht]
\includegraphics[width=\linewidth]{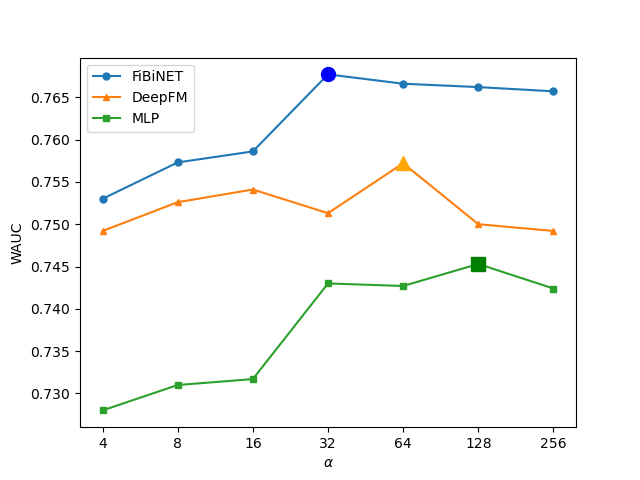}
\caption{
As temperature coefficient \textbf{$\alpha$} changes, the performance of MLP, FiBiNET, and DeepFM with MLoRA varies on the Taobao-10 dataset.}
\label{fig:lora-reduce}
\end{figure}
\textbf{Evaluation of $r$.} 
Note that $r$ is able to influence the performance of MLoRA. 
Thus, we conduct ablation experiments by varying temperature coefficient $\alpha$ to adjust the $r$ value. 
Same as before, we take MLP, FiBiNET, and DeepFM as examples. The ablational results are shown in Figure~\ref{fig:lora-reduce}. 
It can be observed that the performance of all three methods follows a trend of first increasing and then decreasing as $\alpha$ is \ghn{increases}, which indicates the existence of an optimal $\alpha$. 
Additionally, the optimal $r$ value corresponding to different methods varies. 
Specifically, MLP, DeepFM, and FiBiNET achieve optimal performance at $\alpha$ values of 128, 64, and 32, respectively. 
This suggests that more complex models require larger $r$ values, which in turn implies smaller $\alpha$ values.

\textbf{Evaluation of domain size.} To analyze the performance of MLoRA with varying numbers of domains, we conduct ablation experiments using three multi-domain datasets: Taobao-10, Taobao-20, and Taobao-30.
From Table~\ref{table:Ablation-taobao}, it can be observed that MLoRA achieves performance improvements of 0.49$\%$, 0.35$\%$, and 0.29$\%$ on Taobao-10, Taobao-20, and Taobao-30 datasets, respectively.
MLoRA exhibits improvements across datasets with different numbers of domains.
We also note that the improvements are more pronounced when the number of domains is smaller, while they decrease slightly as the number of domains increases. 
\gdh{The reason may be that when the number of domains increases, the base model is more influenced by different domain data, which results in more fluctuations in our MLoRA performance. 
As the base model is frozen in the finetuning phase,   one possible solution is to finetune the base model and LoRA network simultaneously in the future.
}

\begin{table}
\centering
\caption{
The WAUC($\%$) results illustrate the impact of changing the number of neurons in hidden layers on the model's performance. Specifically, "1x," "2x," "3x," "4x," and "5x" correspond to the number of neurons in three-layer FCNs, scaled by [256, 128, 64], [512, 256, 128], [1024, 512, 256], [1024, 1024, 512] and [2048, 1024, 512], respectively. }
\label{table:mlora-para}
\begin{tabular}{c|c|c|c|c|c}
\toprule
\textbf{MLoRA} & 1x & 2x & 3x & 4x & 5x \\
\midrule
\textbf{WAUC} & 75.39 & 75.82 & 75.99 & 76.68 & 76.93 \\
\bottomrule
\end{tabular}
\end{table}

\textbf{Evaluation of Parameter Size of Pretrained Model.} Due to the larger scale of Amazon-6 dataset, we hypothesize that the current three-layer fully connected neural network with hidden layer neuron quantities of (256,128,64) might not be sufficient to learn the data distribution knowledge in large-scale data. 
Therefore, we employ MLP+MLoRA approach and explore the impacts of changing the number of neurons on the model's performance. 
As shown in Table~\ref{table:mlora-para}, it can be observed that the model's performance continuously improves along with the number of neurons in hidden layers increasing.
This demonstrates that larger-scale data requires more complex and parameter-rich networks to enable the model to learn sufficiently.


\subsection{Industry Application}\label{Sec:Industry_App}
\yzm{Our MLoRA approach is already deployed on the Alibaba.COM e-commerce website and an online A/B testing was conducted to further verify the effectiveness.}
We integrated 10 core recommendation domains and utilized data from nearly 90 days (13 billion samples) for pretraining, and data from nearly 21 days (3.2 billion samples) for finetuning to build the MLoRA approach. 
It is a challenging endeavor to deploy the MLoRA approach in a large-scale recommendation system as it serves millions of users daily. 
Considering this, we only deployed the best offline method as our baseline. 
The temperature coefficient $\alpha$ of MLoRA is set to 16 by default, which means that the $r$ value is one-sixteenth of the output dimension size, and the $r$ value has a minimum value of one. 
To prevent the bottom layer parameter size of MLoRA from being too large, we increase the temperature coefficient until the parameter size of $\mathbf{A}$ and $\mathbf{B}$ does not exceed 1024. 
Ultimately, additional parameter size is only increased by $1.76\%$.

We carefully conducted online A/B testing from February to March 2024.
MLoRA contributes to 1.49\% and 3.37\% increases in CTR and in the order conversion rate for recommendation scenarios, as well as a 2.71\% increase in the number of paid buyers across the entire site.
This indicates a significant improvement in proving the effectiveness of MLoRA in a real-world production environment. 
MLoRA has been launched online serving millions of users every day. \textbf{Note that, MLoRA is flexible to newly added domains.} When there is a new domain, adding and finetuning a new LoRA adaptor is very convenient.

\begin{table}
	\centering
	\caption{The CTR, CVR and PB gains in the online recommendation system. CVR refers to the order conversion rate for recommendation scenarios. PB refers to the number of paid buyers across the entire site.}
 \label{tab:Ind_app}
	\begin{tabular}{c|c|c|c}
		\toprule  
		&CTR&CVR&PB\\ 
		\midrule  
		$\Delta$&+1.49\%&+3.37\%&+2.71\%\\
		\bottomrule  
	\end{tabular}
\end{table}

\section{Conclusion}
In this paper, we have proposed a model-agnostic framework (i.e., MLoRA) for multi-domain CTR prediction.
MLoRA alleviates the challenges of data sparsity and insufficient learning by configuring LoRA adaptors for each domain, while also utilizing fewer model parameters. 
Our proposed MLoRA approach is a general framework, which can be applied to various CTR models conveniently.
\yzm{Experimental results on public datasets and online A/B testing results in the Alibaba.COM e-commerce website demonstrate the superiority and feasibility of MLoRA.}

In the future, we \ghn{will} further refine the cooperation of the base model and LoRA networks to accommodate complex issues of multi-domain CTR prediction better as discussed in the evaluation of size in Section~\ref{sec::ablation}. 
Meanwhile, we will further enhance the MLoRA approach to address the challenges associated with data sparsity and inadequate learning capabilities.

\clearpage

\bibliographystyle{ACM-Reference-Format}
\bibliography{main}


\end{document}